\documentclass{IEEEtran}

\usepackage{cite}
\usepackage[utf8]{inputenc}
\usepackage{amsmath,amssymb,amsfonts}
\usepackage{amsthm}
\usepackage{algorithmic}
\usepackage{graphicx}
\usepackage{textcomp}
\usepackage{xcolor}
\usepackage{booktabs}
\usepackage{subcaption}
\newtheorem{theorem}{Theorem}

\def\BibTeX{{\rm B\kern-.05em{\sc i\kern-.025em b}\kern-.08em
    T\kern-.1667em\lower.7ex\hbox{E}\kern-.125emX}}

\begin{document}

\title{UAV-based Maritime Communications: Relaying to Enhance the Link Quality}
\author{\IEEEauthorblockN{Abdullah Taha Çağan$^{1,2}$, Görkem  Berkay Koç$^{1,2}$, Handan Yakın$^{1,2}$,  Berk Çiloğlu$^{1,2}$, Muhammad Zeeshan Ashgar$^2$, Özgün Ersoy$^1$, Jyri Hämäläinen$^2$, Metin Öztürk$^{1,3}$}
\IEEEauthorblockA{\\$^1$Electrical and Electronics Engineering, Ankara Yıldırım Beyazıt University, Ankara, Türkiye.
\IEEEauthorblockA{\\$^2$School of Electrical Engineering, Aalto University, Espoo, Finland.}
\IEEEauthorblockA{\\$^3$Non-Terrestrial Networks Lab, Systems and Computer Engineering, Carleton University, Ottawa, Canada.}}}
\maketitle

\begin{abstract}
Providing stable connectivity in maritime communications is of utmost importance to unleash the full potential of smart ports.
Nonetheless, due to the crowded nature of harbor environments, it is likely that some ships are shadowed by others, resulting in reduced received power that subsequently diminishes their data rates---even threatens basic connectivity requirements.
Given that uncrewed aerial vehicles (UAVs) have been regarded as an integral part of future generations of wireless communication networks, they can be employed in maritime communications as well.
In this paper, we investigate the use of UAV-mounted relays in order to help mitigate the reduced data rates of blocked links in maritime communications.
Various communication architectures are considered based on the positioning mechanism of the UAV; in this regard, fixed, $k$-means algorithm-based, and landing spot-based positioning approaches are examined.
Additionally, since UAVs are predominantly battery-operated, the energy consumption performances of these approaches are also measured. 
Results reveal that the landing spot-based UAV relay positioning approach finds the best trade-off between the data rate and energy consumption.
\end{abstract}


\section{Introduction}
\label{sec:introduction}

There is no shred of doubt that the maritime sector has been experiencing growing demand, encompassing various fields such as transportation, fishing, and recreational boating, to name a few. 
In particular, maritime transportation, which plays a significant role in the economy---also known as the \emph{blue economy}---, accounts for around 80\% of worldwide trade~\cite{psaraftis2021future}. 
On the one hand, approximately 80\% of ports around the world still process their maritime services using manual and conventional operations~\cite{port_paperbased}, while, on the other hand, recent technological developments, including the Internet of Things~(IoT) and the fifth-generation wireless communication networks~(5G), have paved the way for the concept of \textit{smart ports} under the umbrella of Industry 4.0, which promises digitization everywhere~\cite{industry40}. Therefore, it is expected that the near future will bring enormous data traffic around ports due to these digitization processes (i.e., smart ports) and the increasing pervasiveness of IoT technology~\cite{smartport_iot}.
Additionally, in June 2023, the International Telecommunications Union~(ITU) revealed its vision for the sixth-generation wireless communication networks~(6G), which includes \emph{ubiquitous connectivity} as a usage scenario and \emph{connecting the unconnected} as one of the four design principles~\cite{itu_vision_june_23}. This vision aligns with the requirements of smart ports enabled by IoT technology, as all the benefits are realized through the communication of various devices (e.g., sensors, user terminals, servers, etc.).

It is quite possible that there could be numerous ships commuting within the same vicinity in a maritime environment, which may pose a threat to smaller ships that are shadowed by larger ones. For instance, if a larger ship (referred to as a \emph{blocking ship} hereafter) is positioned between a terrestrial base station (BS) and a smaller ship (referred to as a \emph{victim ship} hereafter), the received signal strength of the victim ship may be adversely affected, which, in turn, results in either a lower quality-of-service (QoS) or a complete connection drop. 
In other words, the basic requirements of the smart port concept are violated that would cause a chain of negative impacts starting from local port management issues to economic burdens on a larger scale. 
This indicates that conventional maritime communication systems are not flexible and capable enough to provide smart ports with their distinctive requirements. Besides, large propagation delays, substantial expenses for executing satellite communication, and the lack of bandwidth for MF/HF/VHF communications are among the challenges that are on the table for traditional maritime communication systems~\cite{UAVmaritime}. 
In order to meet the requirements of smart ports as well as realizing the 6G vision of ITU, non-terrestrial networks (NTN) should be utilized in addition to conventional terrestrial communication systems~\cite{NTN}. 

We argue that a customizable relay network is necessary in order to tackle the above-mentioned connectivity-related issues. 
In this regard, uncrewed aerial vehicles (UAVs) as an NTN element would be a viable solution due to their flexibility and mobility characteristics~\cite{flexible_uav}.
More specifically, UAVs are able to present the required feasibility by avoiding the shadowing and reflection effects; because they are airborne, they can adjust their altitude to provide a line-of-sight (LoS) link between the transmitter and receiver.
Considering this LoS advantage, using a UAV relay brings better signal quality compared to the traditional terrestrial relays~\cite{UAVcom}. 
Therefore, to mitigate the disadvantages of conventional maritime communication methods, a UAV-assisted relay can be used. 
Despite their numerous advantages, such as flexibility, mobility, etc., UAVs also have crucial limitations~\cite{flexible_uav}; for example, they are limited by their battery lifetime (for a typical quadcopter, it is approximately 15-30 minutes~\cite{hovering_time}), and thus energy consumption plays a significant role in UAV missions. 

There are two commonly used relaying techniques; namely, decode-and-forward (DF) and amplify-and-forward (AF). 
While the former decodes, re-modulates, and re-transmits the received signal, the latter simply amplifies and re-transmits the signal without decoding ~\cite{levin2012amplify}.  
In this paper, using a UAV-assisted DF relaying system, we aim to provide a better communication performance for the ships with blocked/shadowed links (i.e., victim ships) that receive poor signal strengths. 
In particular, a UAV-mounted relay is placed between the transmitter and receiver, and a multi-hop communication link is established.

\subsection{Literature Review}\label{literature}

There are several works on blocked sea-to-land links, relaying networks for maritime communications, UAV-assisted relay systems, and energy-efficient UAV communications. 
In~\cite{ships}, the authors studied a sea-to-land non-LoS (NLoS) channel, where different types of cargo ships block the propagation path between the transmitter and receiver and make the establishment of a communication link almost impossible.
The channel characteristics for both air-to-sea and
near-sea-surface channel links were investigated in~\cite{wireless} by presenting a comprehensive tutorial for maritime communications.
The work in~\cite{UAV2023} explored the integration of UAVs into maritime communication networks to address the limitations of current maritime communication systems, which include low bandwidth and high costs.
A partial DF (P-DF) cooperative relaying network (CRN) with non-orthogonal multiple access~(NOMA) was proposed in~\cite{P-DF} for maritime communications using a relay deployed on an island.
The study in~\cite{UAVmaritime} considered the caching through a UAV-aided DF relay system in downlink maritime communications, and in order to increase the average achievable rate for a specific number of users, the authors developed an optimization problem for the placement of UAVs. 
The authors in~\cite{overthesea} studied three UAV-aided network designs: \textit{i)} flying remote radio head (RRH) that is similar to the cloud radio access networks (C-RAN); \textit{ii)} flying relays; and \textit{iii)} flying BS. 
They considered the challenges and advantages of the offered architectures from the perspective of maritime search and rescue operations and conducted performance comparison using energy efficiency, data rate, and latency metrics.

To provide fair and reliable communication service while maximizing energy efficiency, the work in~\cite{energy} constructed comprehensive models for UAV-based communication systems, including channel, data rate, and energy models, and discussed the use of deep reinforcement learning to create an energy-efficient control strategy for 3D UAV control.
The research in~\cite{multirelay} utilized network service-oriented multi-UAV-assisted relays between ships and fixed onshore BS via $Q$-learning to stabilize QoS. 
To improve the effectiveness of relay communication channels, UAVs might pre-plan their flight trajectories in accordance with their past experiences.
To provide a container vessel ship with data communication and mitigate transferring delay, the authors in~\cite{offshore} proposed a DF-based UAV-aided relay from the container vessel ship to an onshore BS. 
It takes into account the port entrance mobility of the container vessel ship, the movement of UAV, and the parameters of marine wireless transmission.

The study in~\cite{coverage} integrated fixed-wing UAVs into the satellite-terrestrial communication networks in a maritime scenario to enhance coverage. 
Satellites and UAVs share the same spectrum, and UAVs use satellites or onshore BSs for wireless backhauling.
The authors in~\cite{bouy} considered a single offshore UAV as a DF relay in the scenario of offshore buoy communication, where the UAV enhances service for energy-scarce ocean buoy networks by serving as both a flight power station and an information relay node.
The research in~\cite{Eff_tra_plan} discussed optimizing the trajectory of UAVs to facilitate data transmission from vessel users to onshore BSs. The authors presented an algorithm that jointly optimizes the UAV trajectory and user scheduling by leveraging the successive convex approximation technique, thereby aiming to provide a satisfactory data transmission for each user, minimizing the energy consumption of the UAV.

Using UAV swarms and NOMA methodologies, a comprehensive approach to improving the connectivity of maritime communication networks was presented in~\cite{UAV_swarm}. 
For uplink and downlink communications, the authors proposed buffer-aided and UAV-based NOMA techniques to maximize the sum rate without requiring complicated hardware or extensive channel state information at the shore BS.
The authors in~\cite{wang2024trajectory} studied optimizing the trajectory of UAVs in maritime communications to improve energy efficiency. They introduced a new clustering algorithm involving NOMA along with wireless power communication (WPC) technology for charging marine terminals. The results demonstrated enhancements in throughput performance and energy efficiency.
The above literature review is summarized in Table~\ref{tab:my-table}.
\begin{table*}[]
\centering
\caption{Literature Review}
\label{tab:my-table}
\begin{tabular}{cllll}\toprule
\textbf{REF.}                & \textbf{YEAR} & \textbf{METHODS}                                                                                                                                                                       & \textbf{OBJECTIVE}                                                                                                                        & \textbf{OPTIMIZATION VARIABLE}                                                                                   \\ \hline
~\cite{ships}  &2011 &\begin{tabular}[c]{@{}l@{}}  Sea-to-land NLoS sea channel using a hybrid\\ of the shooting bouncing ray (SBR)
algorithm and \\ the uniform theory of diffraction (UTD).\end{tabular}                                                                                                                                                   & \begin{tabular}[c]{@{}l@{}}Characterization of \\ the sea-to-land NLOS channel.\end{tabular}                                     & Stand-off distance                                                                                                        \\ \hline
~\cite{wireless}    & 2018 & \begin{tabular}[c]{@{}l@{}} Channel characteristics
for the air-to-sea and\\ near-sea-surface channel links.\end{tabular}                                                                                                                                                                            & Investigation of the channel links.                                                                                                  & User distribution                                                                                                    \\ \hline
~\cite{UAV2023} & 2023 & \begin{tabular}[c]{@{}l@{}} Comprehensive review and analysis \\ approach to explore the integration of UAVs \\ into maritime communication networks.\end{tabular} & \begin{tabular}[c]{@{}l@{}} Improve connectivity and efficiency.\end{tabular}                                                                                              & \begin{tabular}[c]{@{}l@{}}UAV trajectory optimization\end{tabular}                          \\ \hline

~\cite{P-DF}        & 2020 & \begin{tabular}[c]{@{}l@{}}P-DF-CRN-NOMA for maritime communications,\\ the maximum ratio
combining (MRC) is utilized.\end{tabular}                                                                   & Effectiveness of proposed protocol.                                                                                              & User mobility                                                                                           \\ \hline
~\cite{UAVmaritime} & 2020 & \begin{tabular}[c]{@{}l@{}}UAV-assisted DF in a downlink maritime \\communication by applying one-dimensional \\linear search optimization.\end{tabular}                                                      & \begin{tabular}[c]{@{}l@{}}Caching through UAV,\\ Maximization the average achievable rate.\end{tabular}                             & Placement of UAV                                                                            \\ \hline
~\cite{overthesea}  & 2019 & \begin{tabular}[c]{@{}l@{}}Three network architecture designs using UAV-\\aided networks for maritime communications.\end{tabular}                                                        & \begin{tabular}[c]{@{}l@{}}Comparing three networks design in term of \\ data rate, latency and energy efficiency.\end{tabular}    & \begin{tabular}[c]{@{}l@{}}UAV's location \\and payload weight\end{tabular}  \\ \hline
~\cite{energy}      & 2020 & \begin{tabular}[c]{@{}l@{}}3-D UAV deployment scheduling algorithm based \\ on deep deterministic policy gradient algorithm.\end{tabular}                                     & \begin{tabular}[c]{@{}l@{}}Maximization of energy efficiency \\ to provide reliable communication.\end{tabular}                       & Energy of UAV                                                                                                      \\ \hline
~\cite{multirelay}  & 2021 & \begin{tabular}[c]{@{}l@{}}Network service-oriented multi-UAV-assisted relays\\ between ships and fixed onshore BS via $Q$-learning.\end{tabular}                               & \begin{tabular}[c]{@{}l@{}}Data rate configuration \\ and marine environment adaptability.\end{tabular}                           & UAV trajectory                                                                                         \\ \hline
~\cite{offshore}    & 2021 & \begin{tabular}[c]{@{}l@{}}UAV-aided relay by DF based on\\only direct access (ODA) algorithm and\\ random direct access (RDA) algorithm.\end{tabular}                              & \begin{tabular}[c]{@{}l@{}}Providing data communication for container \\ vessel ship and mitigate transferring delay.\end{tabular} & \begin{tabular}[c]{@{}l@{}}Channel conditions, container \\ data size and available energy\end{tabular} \\ \hline
~\cite{coverage}    & 2020 & \begin{tabular}[c]{@{}l@{}}UAVs into satellite-terrestrial link networks via \\successive convex optimization (SCO)\\ and bisection searching tools.\end{tabular}               & \begin{tabular}[c]{@{}l@{}}A hybrid satellite-terrestrial \\ network's coverage expansion.\end{tabular}                          & \begin{tabular}[c]{@{}l@{}}UAV trajectory, \\ in-flight transmit power\end{tabular}                    \\ \hline
~\cite{bouy}        & 2020 & \begin{tabular}[c]{@{}l@{}}Offshore UAV as a relay by DF in buoy\\communication  by applying the block coordinate\\ descent and successive convex optimization.\end{tabular} & Ensuring the communication quality.                                                                                              & \begin{tabular}[c]{@{}l@{}}UAV trajectory, \\ transmission power\end{tabular}                          \\ \hline
~\cite{Eff_tra_plan}        & 2023 & \begin{tabular}[c]{@{}l@{}}Successive convex approximation\\technique is executed.\end{tabular} & \begin{tabular}[c]{@{}l@{}}Minimize UAV energy consumption,\\ensure data transmission for individual users.\end{tabular}                                                                            & \begin{tabular}[c]{@{}l@{}}UAV trajectory, user scheduling\end{tabular}                          \\ \hline
~\cite{UAV_swarm}        & 2024 & \begin{tabular}[c]{@{}l@{}}Buffer-aided NOMA technique is applied \\in downlink and uplink sides.\end{tabular} & \begin{tabular}[c]{@{}l@{}}Increase the data rate of the network,\\ improve spectral efficiency, reduce latency.     \end{tabular}                                                                                           & \begin{tabular}[c]{@{}l@{}}Power allocation coefficients\end{tabular}                          \\ \hline
~\cite{wang2024trajectory}   & 2024 & \begin{tabular}[c]{@{}l@{}}A NOMA-based UAV-assisted marine\\ information collection system using\\ dynamic clustering and convex optimization\\ algorithms to enhance energy efficiency.\end{tabular} & Prolong maritime communications energy.      & \begin{tabular}[c]{@{}l@{}} UAV trajectory,\\ charging and time allocation\end{tabular}                          \\ \hline
\end{tabular}
\end{table*}
\subsection{Motivation and Contribution} \label{contributions}
Maritime communication is a comprehensive field in which various studies can be accomplished. 
However, to the best of our knowledge, UAV-enabled blocking avoidance and the energy consumption of UAV-aided relays for different UAV placement scenarios in maritime communications have merely been investigated in the literature. These gaps motivate us to examine NLoS situations in maritime applications and improve communication performance using a UAV-aided DF relay positioned for optimal energy efficiency. 
This work's key contributions are summarized as follows:
\begin{itemize}
    \item We propose a UAV-assisted DF relaying system for maritime communications to enhance the link quality for victim ships.
    \item Four different communication architectures for non-UAV (without the use of a UAV relay) and UAV-aided DF relay systems are considered. To implement these communication architectures, we evaluate three different UAV placements to achieve the maximum data rate in both single-victim-ship and multiple-victim-ship scenarios.
    \item A mathematical model showcasing the relationship between the data rate and the distance of the UAV relay is derived to establish the underlying hypothesis of the work: \textit{the data rate of the victim ships can be boosted by optimizing the positioning of the UAV relay}.
    \item We investigate the landing spot~(LS) approach, a novel energy-efficient UAV placement methodology where UAVs perch on predesignated spots (e.g., rooftops, lamp posts, etc.) instead of hovering in the air, to analyze its impacts on both energy efficiency and data rate.
\end{itemize}
\subsection{Organization of the Paper}\label{organisations}

The rest of this paper is organized as follows: the system model, considered communication architectures, and the system scenarios are presented in Section~\ref{sec:systemmodel}. 
The problem formulation is presented in Section~\ref{sec:problemformulation}, while Section~\ref{sec:performanceevaluation}  includes the simulation parameters, numerical results and discussions. Lastly, Section~\ref{sec:conclusion} concludes the paper. 

\section{System Model} \label{sec:systemmodel}

\begin{figure*}[t]
\begin{subfigure}{0.48\textwidth}
    \includegraphics[width=\linewidth]{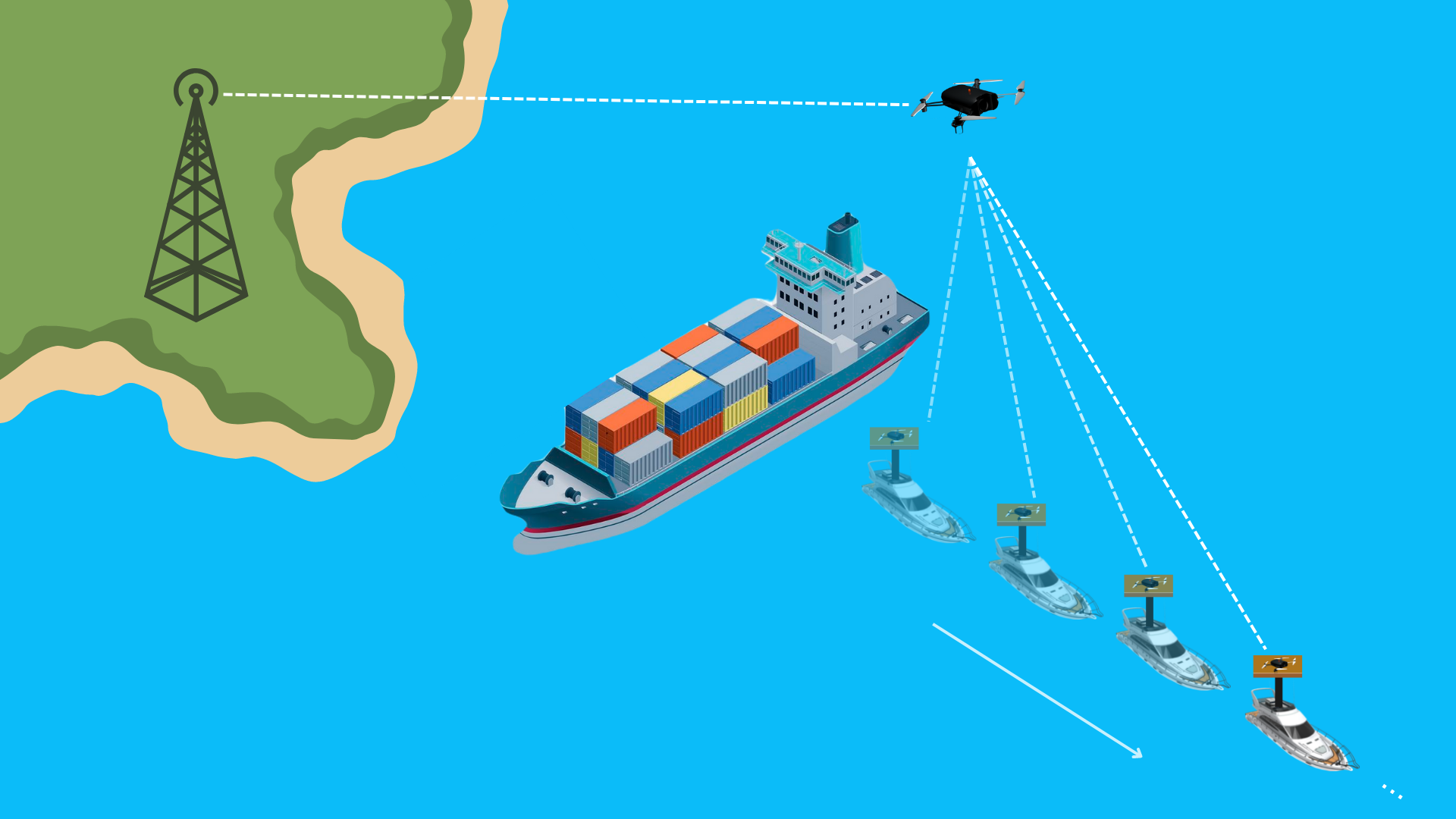}
    \caption{Illustration of the single-victim-ship scenario.}
    \label{fig1}
\end{subfigure}
\hfill
\begin{subfigure}{0.48\textwidth}
    \includegraphics[width=\linewidth]{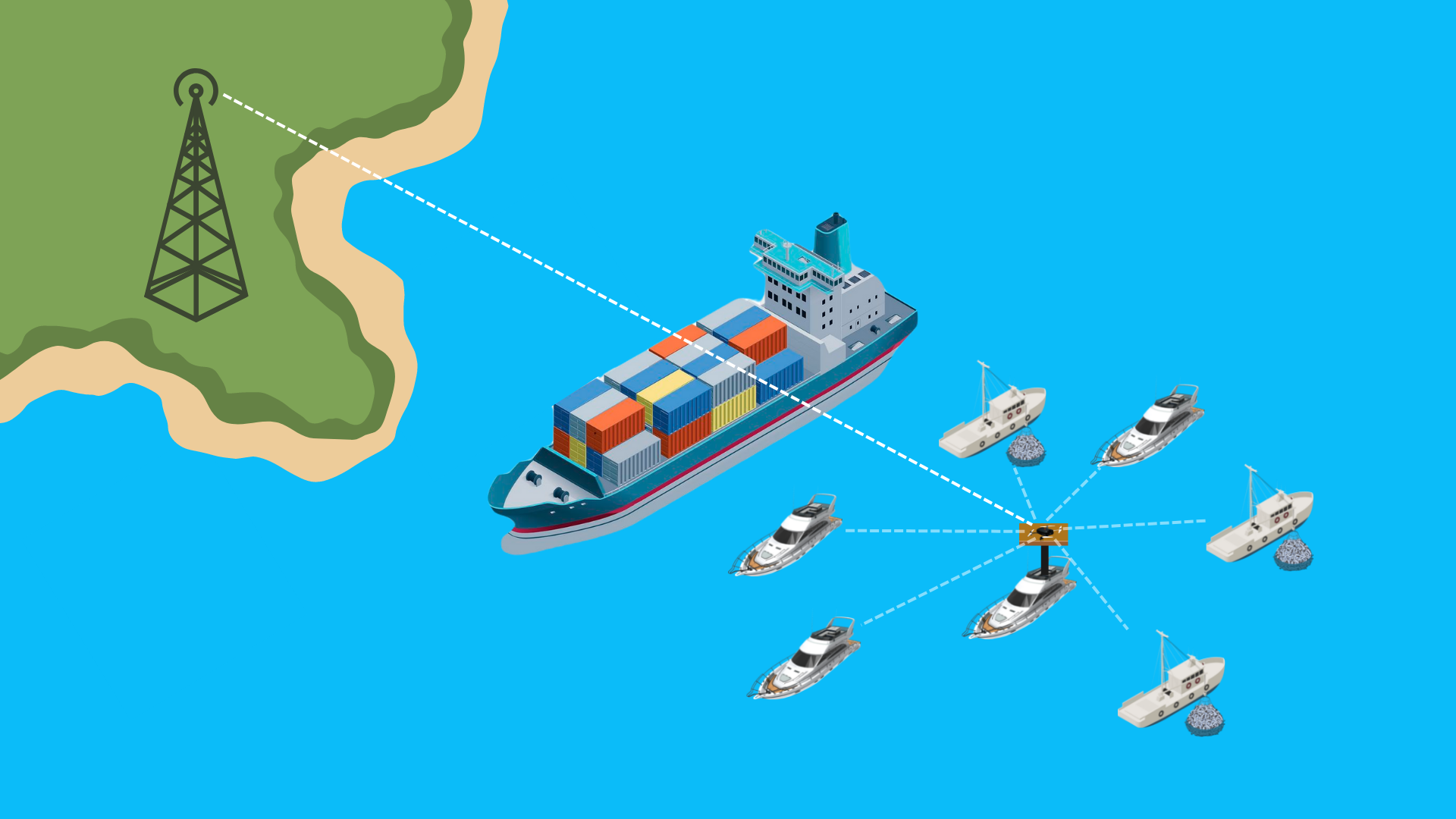}
    \caption{Illustration of the multiple-victim-ship scenario.}
\label{fig2}
\end{subfigure}
\caption{Illustration of the system scenarios.}\label{figure1}
\end{figure*}

In this work, we consider a UAV-assisted DF relaying system for maritime communications. The network consists of a single onshore (terrestrial) BS, a UAV relay, one or more victim ships, and a blocking ship. The blocking and victim ships are positioned within the footprint of the terrestrial BS, as illustrated in Fig.~\ref{figure1}. 
Next, we explore different positioning approaches (referred to as \textit{communication architectures}, which will be elaborated on in Section~\ref{Relaying_Situation}) for the UAV relay that vary according to the movement of the victim ship(s). It should be noted that the blocking ship remains stationary in all scenarios.

\subsection{Communication Architectures}\label{Relaying_Situation}
 Four different communication architectures are considered in this work; namely, \textit{i)} no-relay (NR), \textit{ii)} fixed-positioned UAV relay (FPR), \textit{iii)} clustering based full-mobile UAV relay (CFMR), and \textit{iv)} LS assisted semi-mobile UAV relay (LSMR).
 They will be detailed in the following paragraphs.

\subsubsection{No Relay (NR)}
In this architecture, no relaying mechanism is included, such that a conventional single-hop wireless communication system is established between the terrestrial BS and the victim ship.

\subsubsection{Fixed Position UAV Relay (FPR)}\label{Relaying_Stationary}
A UAV-mounted DF relay system is positioned between the terrestrial BS and victim ship, and the position of the UAV relay is fixed using two different approaches.
In the first approach, the $x$ and $y$ coordinates of the UAV relay are fixed to the middle point of the initial distance\footnote{Initial distance refers to the distance calculated between the terrestrial BS and victim ship at the first instance---at the zeroth time slot.} between the terrestrial BS and the victim ship. In the second approach, according to the multiple victim ships' movement areas, the $x$ and $y$ coordinates of the UAV relay stay in the middle of the area-of-interest. The $z$ coordinate, on the other hand, is adjusted to the minimum that provides a LoS link.

\subsubsection{Clustering Based Full-Mobile UAV Relay (CFMR)} \label{relayingcasecentroid}
The position of the UAV-mounted DF relay is determined with the $k$-means algorithm; such that the victim ships are considered to be the data points while the ($x,y$) position of the UAV relay becomes the centroid. The number of clusters is set to one as there is only one UAV relay considered\footnote{The number of UAV relays would go up according to the number of victim ships. In multiple UAV relay case, the number of clusters (i.e., the number of UAV relays) can be determined with the elbow method.}, and the UAV hovers at the location of the centroid.

\subsubsection{Landing Spot Assisted Semi-Mobile UAV Relay (LSMR)}
Unlike the CFMR case, the UAV relay does not hover; it rather perches on a predesignated point (e.g., flag pole) called LS~\cite{LandingSpot_Definition}. 
 In the single-victim-ship scenario, the UAV relay's position is fixed to on top of the LS on the victim ship, whereas in the multiple-victim-ship scenario the $k$-means algorithm is employed and the victim ship that is closest\footnote{In terms of the Euclidean distance.} to the centroid is selected as the LS for the UAV relay.
 As victim ships move, the victim ship that hosts the UAV relay is updated with subsequent reruns of $k$-means algorithm.
 
\subsection{System Scenarios}
In order to test the various communication architectures mentioned in Section~\ref{Relaying_Situation} and evaluate the results, two different system scenarios are considered; namely, single-victim-ship and multiple-victim-ship scenarios.

\subsubsection{Single-Victim-Ship Scenario} \label{singlevictimship}
This scenario consists of a terrestrial BS and a single victim ship, where the victim ship suffers from the shadowing effects caused by the blocking ship.
According to the system scenario, the victim ship moves in a specified direction, as seen in Fig. \ref{fig1}. 
It is considered that the victim ship has an LS, where a UAV relay can perch on.
\subsubsection{Multiple-Victim-Ship Scenario}
In this scenario, there are multiple victim ships that require to communicate with the terrestrial BS.
The victim ships are scattered in the area-of-interest randomly with uniform distribution.
Due to the effect of the blocking ship, the QoS is disturbed, and to tackle this problem, a UAV-assisted DF relay is considered.
Like the single-victim-ship scenario, the victim ships are considered to move in specified directions\footnote{This time two different directions, which are opposite to each other and parallel to the shore, are assumed.}.
As a result of the movement of victim ships at different time intervals, the UAV changes its position to provide better communication performance (except for FPR, where the UAV relay is fixed-positioned).


\subsection{Propagation Model}
The terrestrial BS, UAV-aided relay, and vessel nodes operate in half-duplex and single-input single-output (SISO) mode. 
It is assumed that the terrestrial BS, UAV relay and the victim ships use an isotropic antenna. 
For practical cases, it is considered to add power gain to all links (i.e., BS–victim ship, relay–victim ship, BS-relay).  However, since power gains in these links depend on the applied antennas and system geometry (that is, how BS, relay, and victim ship are located) we rather set all gains 0 dBi to avoid situations where antennas and system geometry favor either the direct link or the relayed link.
Since the goal of this research is to provide a maritime communication system that mitigates the large-scale fading caused by the blocking ship, establishing a path loss model becomes crucial. 
It would be advantageous to use a two-ray path loss model in an environment with a dominant LoS link; however, the two-ray path loss model is useless in this case due to the NLoS environment being dominant~\cite{Wimax}.
Hence, the propagation model adopted in this work is given by
\begin{equation} \label{freePL}
    L = 20 \log_{10}\left(\frac{4\pi f_\text{c} d}{c}\right)+\zeta,
\end{equation}
where $f_\text{c}$ is the carrier frequency, $d$ is distance, $c$ is speed of light, and $\zeta$ is additional loss.
For convenience, the effects of sea wave motion and ducting effects are neglected.
In the NLoS situation, the received power may drop approximately by 10-20 dB~\cite{Wimax}, which is captured by $\zeta$ in~\eqref{freePL}.

In a SISO channel, the signal-to-noise ratio (SNR) between the BS and the victim ship is determined using

\begin{equation}\label{snrbstoship}
    \operatorname{SNR}_\text{SISO} = \frac{P_\text{s,t}|h_\text{bv}|^2}{\sigma^2},
\end{equation}
where $P_\text{s,t}$ is the BS to victim ship transmit power, $h_\text{bv}$ shows the deterministic flat-fading channel between the BS and the victim 
ship~\cite{Emil}.
The data rate between the BS and the victim ship without any relaying mechanism (i.e., SISO mode) is given by

\begin{equation}\label{noUAV}
    R_\text{SISO} = \log_2\left(1 + \operatorname{SNR}_\text{SISO}\right).
\end{equation}

The relay uses a DF relaying protocol\footnote{For the details about the DF relaying adopted in this paper, refer to~\cite{Emil}.}.
The achievable rate at the victim ship by employing a DF relaying mechanism can be computed by
\begin{equation}\label{rate}
    R_\text{DF} = \frac{1}{2} \log_2\left(1 + \operatorname{SNR}_\text{DF}\right), 
\end{equation} 
where $\operatorname{SNR}_\text{DF}$ is the SNR value obtained at the receiver as
\begin{equation}
    \operatorname{SNR}_\text{DF} = \min\left(\frac{P_\text{s,t} |h_\text{br}|^2}{\sigma^2},\frac{P_\text{s,t}|h_\text{bv}|^2}{\sigma^2}+\frac{P_\text{u,t} |h_\text{rv}|^2}{\sigma^2}\right),
\end{equation}
where $h_\text{br}$ and $h_\text{rv}$  are the channel between BS to UAV, and UAV to victim ship, respectively. $P_\text{u,t}$ is UAV to victim ship transmit power.

Since there is a relay between the victim ship and the BS, the signal takes two different hops, doubling the required transmission time. The transmission of the signal initially takes place between the BS and the relay, and then the transmission of the same symbols must occur in a second stage between the relay and the victim ship.
In this regard, BS-to-relay SNR value is $P_\text{s,t}|h_\text{br}|^2/\sigma^2$, BS-victim ship SNR value is $P_\text{s,t}|h_\text{rv}|^2/\sigma ^2$, and the relay-victim ship SNR value can be expressed as $P_\text{s,t}|h_\text{bv}|^2/\sigma^2$.
 Equation (5) indicates that in case the channel quality of one of the hops is worse than the other because the symbols received in the two phases are the same, the signal quality of the worst will determine the available rate.
In addition, when calculating the SNR of the victim ship, the BS-victim ship link with NLoS should be taken into account, as well as the relay-victim ship link with LoS.

\subsection{Energy Consumption Model} \label{subsection: energycons}
The total energy consumption of the UAV can be modeled as the combination of propulsion energy and communication energy. 
The latter is consumed by the communication circuitry, transmitting/receiving signals between ships and BS, while the former is required to hover and support the mobility (taking-off, landing, etc.) of the UAV.

In this paper, because we use a fixed transmit power at the BS and relay, it is assumed that the communication energy is constant for all the scenarios. On the other hand, the propulsion energy varies according to the mobility/hover status of the UAV. The energy consumption parameters are obtained from~\cite{energy} and given in Table \ref{table_energy}.

\begin{table}[ht]
\centering
\caption{Energy Consumption Parameters} \label{table_energy}
\resizebox{\columnwidth}{!}
{
\begin{tabular}{lll}
\textbf{Notations} & \textbf{Parameters}        & \textbf{Value} \\ \hline
$N$                  & Number of rotors            &         4       \\
$V$                   & Thrust                     &    34.3 N           \\
$W$                   & Weight of frame            &       1.5 kg         \\
$m$                   & Weight of battery and payload    &    2 kg            \\
$g$                   & Gravity                    &        9.8 N/kg           \\ 
$r$                   & Rotor disk radius          &  0.4              \\
$\rho$                   & Fluid density of the air   &  1.225 $\text k\text g/\text m^3$           \\
$P_\text{u,t}$                 & Relay transmission power       &     0.0316 W           \\
$P_\text{cu}$                & Onboard circuit power       &    0.01 W            \\
$C_x$                   & Drag coefficient           &       0.025         \\
$A$                   & Reference area             &        0.192 $\text m^2$        \\
$c_b$                   & Rotor chord                &     0.022 m           \\
$w$                     & Angular velocity            & 16 rad/sec \\
$v_\text{a}$                   & Up speed of UAV             & 10 m/s \\
$v_\text{d}$                   & Down speed of UAV           & 10 m/s \\
$v_\text{h}$                   & Horizontal speed of UAV (LSMR)     & 27.7 m/s \\
$v_c$                   & Horizontal speed of UAV (CFMR) & 10 m/s \\     
\bottomrule
\end{tabular} 
}
\end{table}

 \subsubsection{Communication Energy Consumption}
The communication energy is needed to establish a communication link between the BS and victim ships.
The communication energy consumption of the UAV can be given by~\cite{energy}
\begin{equation}\label{eq:comenergy}
    E_\text{C} (t_j) = \left(n_\text{u} P_\text{u,t} + P _\text{cu}\right)t_\text{com},
\end{equation}
where $P_\text{cu}$ is the onboard circuit power (in Watts) and $P_\text{u,t}$ is the communication-related power (in Watts). 
$t_\text{com}$ is the duration of communication, and $n_\text{u}$ is the number of the served victim ships by the UAV at time slot $t_j$.

\subsubsection{Hovering Energy Consumption}
The hovering energy is required to keep the UAV at a certain altitude, and is calculated by~\cite{energy}
\begin{equation} \label{eq:hover_powercons}
    P_\text{hv} = \frac{N V^\frac{3}{2}}{\sqrt{2\rho\pi r^2}},
\end{equation}
where $N$ is the number of rotors of the UAV, and $V=(W+m) g$ is the thrust, where $W$ is the weight of the frame, $m$ is the battery and payload weight, and $g$ is the gravity. $\rho$ is the fluid density of the air and $r$ is the rotor disk radius.

Then, the hovering energy consumed by the UAV at $t_j$ is obtained by
\begin{equation}\label{eq:hover_energycons}
    E_\text{H} (t_j) = P_\text{hv}t_\text{hv},
\end{equation}
where $t_\text{hv}$ is the duration of hovering.

\subsubsection{Mobility Energy Consumption}
Mobility energy is the energy consumed by the UAV while moving from one point to another, and is calculated as~\cite{energy} 
\begin{equation} \label{eq:mobil_energycons}
\begin{split}
    E_\text{M} (t_j) = P_\text{h} \frac{d(t_j)}{v_\text{h}} + I \left(\Delta h(t_j)\right) P_\text{a} \frac{\Delta h(t_j)}{v_\text{a}} \\- \left(1- I \left(\Delta h(t_j)\right)\right) P_\text{d}\frac{\Delta h(t_j)}{v_\text{d}},
\end{split}
\end{equation}
where $P_\text{h}$ is instantaneous power consumption in the horizontal direction, $P_\text{a}$ is ascending power, and $P_\text{d}$ is descending power. $d(t_j)$ is the horizontal moving distance at $t_j$, $\Delta h(t_j)$ is the variety in the altitude, $v_\text{h}$ is horizontal, $v_\text{a}$ is ascending, and $v_\text{d}$ is descending velocities.
The indicator function, $I(\Delta h(t_j))$, is 1 when $ \Delta h(t_j) \geq 0$, and 0 otherwise.
The calculations of power consumption values (i.e., $P_\text{h}$, $P_\text{a}$, $P_\text{d}$) can be found in~\cite{energy} (Eqns. 18-21 therein).

\section{Problem Formulation} \label{sec:problemformulation}
This study aims to enhance the data rate performance of the communication system under the effect of large-scale fading in communication architectures presented by blocking ships and depending on the positioning mechanism of the UAV relay.
To achieve this goal, before proceeding to the performance evaluation, the relationship between data rate and distance is derived from the path loss model employed. 
The formulation procedure begins with the range equation, which relates the received power to the distance between the terrestrial BS and receiver.
The range equation differs according to the presented communication architectures, and, without loss of generality, the following derivation assumes the NR architecture where there is a single transmitter and single receiver\footnote{The derivation procedure and steps may vary for other communication architectures; however, the conclusion drawn will be the same: the data rate is a function of the distance between the transmitter and receiver.}.
\begin{theorem}
The data rate of the victim ships can be enhanced by optimizing the positioning the UAV relays.    
\end{theorem}
\begin{proof}
Assuming a single BS with a transmit power of $P_\text{t}$ and a receiver with a received power of $P_\text{r}$, if we express the free-space path loss model ($L_0$) given in \eqref{eq:1} linearly, we obtain
\begin{equation} \label{eq:1}
    L_0 = \left(\frac{4\pi d}{\lambda}\right)^2.
\end{equation}

When the receiving antenna is isotropic, the received power, $P_\text{r}$, can be defined by
\begin{equation} \label{eq:2}
    P_\text{r} = \left(\frac{P_\text{t} G_\text{t} G_\text{r} \lambda^2}{(4\pi d)^2}\right) = \frac{P_\text{t} G_\text{t} G_\text{r}}{L_0},
\end{equation}
where $G_\text{t}$ and $G_\text{r}$ represent the gain of the transmitting antenna and receiving antenna, respectively. 
SNR is the primary parameter used to measure the quality of signal performance. The expression for SNR is given as
\begin{equation} \label{eq:6}
    \frac{S}{\mathcal{N}} = \frac{P_\text{r}}{\mathcal{N}} ,
\end{equation}
where $S$ and $\mathcal{N}$ are the signal power and thermal noise power, respectively.
Substituting the thermal noise power, formulated as $\mathcal{N} = k T B$, in~\eqref{eq:6}, the SNR expression becomes
\begin{equation} \label{eq:7}
    \frac{P_\text{r}}{\mathcal{N}} = \frac{S}{k T B}.
\end{equation}
Here, $k$, $T$, and $B$ denote Boltzmann’s constant, temperature, and bandwidth, respectively.
By substituting \eqref{eq:2} in \eqref{eq:7}, $P_\text{r}/\mathcal{N}$ is obtained as
\begin{equation} \label{eq:8}
    \frac{P_\text{r}}{\mathcal{N}} = \frac{P_\text{t} G_\text{t} G_\text{r} /\mathcal{N}}{L_0}.
\end{equation}
By replacing the noise power, $\mathcal{N}$, with the noise power spectral density, $\mathcal{N}_0 = \frac{\mathcal{N}}{B} = k T$, we get
\begin{equation} \label{eq:9}
    \frac{P_\text{r}}{\mathcal{N}_0} = \frac{P_\text{t} G_\text{t} G_\text{r} / T}{k L_0}.
\end{equation}
Then, the $\mathcal{E}_b/\mathcal{N}_0$ expression can be derived using SNR as 
\begin{equation} \label{eq:10}
    \frac{\mathcal{E}_b}{\mathcal{N}_0} = \frac{P_\text{r}}{\mathcal{N}} \left(\frac{B}{R}\right) = \frac{P_\text{r}}{\mathcal{N}_0} \left(\frac{1}{R}\right),
\end{equation}
where $R$ denotes the data rate.
Using (\ref{eq:9}) and \eqref{eq:10} data rate is obtained as
\begin{equation} \label{eq:11}
   R = \frac{P_\text{t} G_\text{t} (G_\text{r} / T) \mathcal{N}_0}{k \mathcal{E}_b L_0}.
\end{equation}

Substituting (\ref{eq:1}) into \eqref{eq:11}, the data rate is obtained as
\begin{equation} \label{eq:12}
   R = \frac{P_\text{t} G_\text{t} (G_\text{r} / T)}{k \left(\frac{4\pi d}{\lambda}\right) \left(\frac{\mathcal{E}_b}{\mathcal{N}_0}\right)},
\end{equation}
which is a function of the distance between the BS and the receiver.
By utilizing the formula presented in~\eqref{eq:12}, it has been shown that the maximum data rate can be achieved while maintaining a minimum distance, which can be obtained by optimizing the UAV relay. 



This means that the $h_\text{br}$ and $h_\text{rv}$ channels are assessed individually to obtain $\operatorname{SNR}_\text{DF}$. In contrast, for the LSMR architecture, the UAV relay is positioned on the LS on the victim ship, and the data rate and $\operatorname{SNR}_\text{DF}$ are calculated for the distances that vary according to the movement of the victim ship.
\end{proof}


\section{Performance Evaluation} \label{sec:performanceevaluation}
This section presents the performance metrics used, followed by a comprehensive discussion of the obtained results. The simulation parameters are given in Table \ref{table1}. 
The simulation campaigns are carried out via the MATLAB simulation program.

\begin{table}[ht]
\centering
\caption{Simulation Parameters}\label{table1}
\resizebox{\columnwidth}{!}{%
\begin{tabular}{ll}
\textbf{Parameters}          & \textbf{Value}             \\ \hline
Length of environment size       & 600 m$\times$800 m                 \\
Number of victim ships       & 20 \\
Landing spot height          & 35 m \\
BS antenna gain    & 0 dBi                       \\
Relay antenna gain           & 0 dBi                       \\
Victim ship antenna gain     & 0 dBi                       \\
Victim ship antenna height   & 2 m    \\
BS height          & 35 m                       \\
Blocking ship height            & 32.3 m~\cite{ships}        \\
Blocking ship width/length    & 32 m/200 m               \\
Victim ship height           & 5 m                        \\
Victim ship width/length   & 4 m/20 m                  \\
BS transmit power  & 45 dBm                     \\
Relay transmit power         & 15 dBm                     \\
Additional loss for NLoS     & 20 dB~\cite{Wimax}                    \\
Additional loss for LoS      & 1 dB~\cite{paulo}               \\
Carrier frequency            & 5.8 GHz                    \\
Bandwidth                    & 10 MHz                      \\
Noise figure                 & 10 dBm                      \\
Receiver reference sensitivity & -94.5 dBm~\cite{reference_sensitivity} \\
Time slot duration                & 10 s \\
Number of time slots in multiple-victim-ship scenario           & 20 
\\ 
Number of time slots in single-victim-ship scenario & 10
\\
\bottomrule  
\end{tabular}
}
\end{table}

\subsection{Performance Metrics}
There are two different performance metrics employed in this paper: namely, data rate and energy consumption. 

\subsubsection{Data Rate}

 In this paper, the data rate metric is captured by the average data rate (denoted by $\overline{R}$) which is the mean of all victim ships' data rate values at each time slot.
 Therefore, it can be calculated by dividing the accumulated data rate of all the victim ships by the number of victim ships.

\subsubsection{Energy Consumption}

The total energy consumption is measured by accumulating the energy consumption values resulting from communication, hovering, and mobility, which are determined by \eqref{eq:comenergy}, \eqref{eq:hover_energycons}, and \eqref{eq:mobil_energycons}, respectively.

\subsection{Results and Discussion}
\begin{figure}[b]
\centering
\includegraphics[width=1\linewidth]{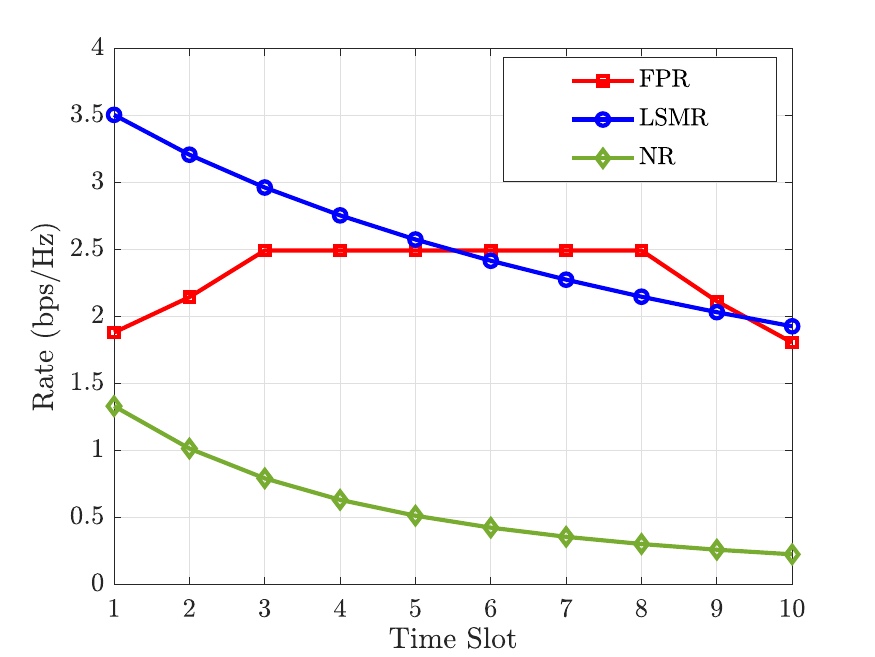}
\caption{Data rates for single-victim-ship scenario.}
\label{fig:singlerate}
\end{figure}

Fig. \ref{fig:singlerate} displays the data rate with respect to time for the single-victim-ship scenario with NR, FPR and LSMR communication architectures.
As can be seen, the data rate of NR is the lowest at all time slots.  
Since the distance between the BS and the UAV increases at each time slot, the data rate decreases in LSMR and NR. 
However, in FPR, the data rate increases until the third time slot as the victim ship gets closer to the UAV. 
For the next five time slots, the data rate remains constant, then it starts decreasing. 
FPR and LSMR offer approximately 3 and 3.5 times better data rates than NR, respectively. 
Thus, the drop due to the shadowing effect is alleviated with the help of the UAV relay.
On the other hand, LMSR provides a 12.66\% better data rate than FPR, because the UAV position is dynamically updated in LSMR, whereas UAV is fixed in FPR. 

\begin{figure}[t!]
\centering
\includegraphics[width=1\linewidth]{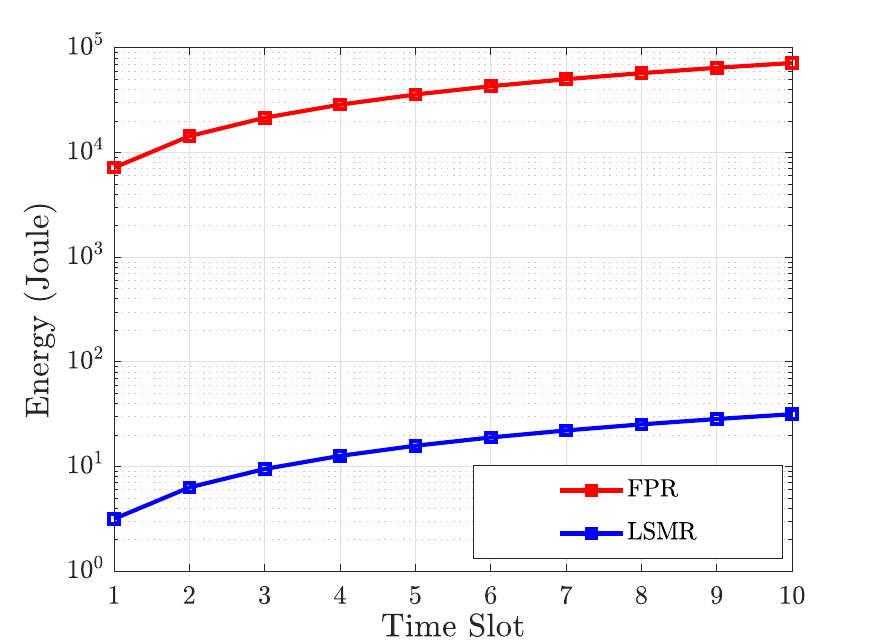}
\caption{Energy consumption for single-victim-ship scenario.}
\label{fig:singleenergy}
\end{figure}

Fig. \ref{fig:singleenergy} presents the energy consumption results for the single-victim-ship scenario. 
Only the results of FPR and LSMR are included in Fig. \ref{fig:singleenergy}, since there is no energy consumption in NR---it does not have a UAV relaying mechanism. 
Because, unlike FPR, the mobility energy consumption is also in place in LSMR, it consumes 74.36\% less energy than FPR. This difference is based on the continuously hovering nature of FPR and the effect of LS strategy that belongs to LSMR.

Having an observation from a dual aspect that combines both data rate and energy behaviors of two communication architectures in the single-victim-ship scenario, the results lead us to infer that LSMR yields the best outcome with higher data rate and lesser energy consumption when it is compared to FPR.

\begin{figure}[b]
    \centering
    \includegraphics[width=1.1\linewidth]{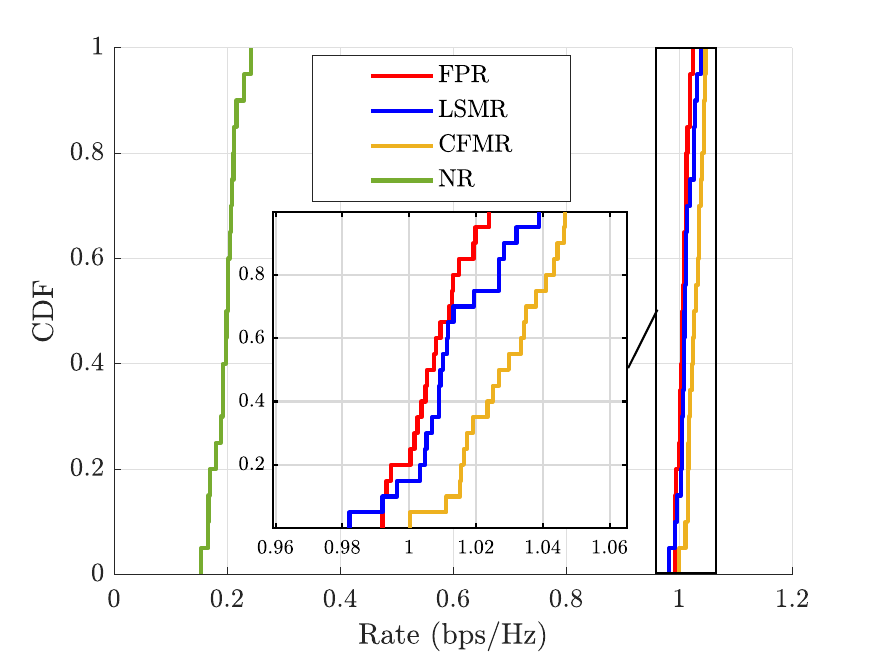}
    \caption{Multiple-victim-ship scenario for all communication architectures.}
    \label{fig:multirates}
\end{figure}


The cumulative distribution function of data rates for multiple-victim-ship scenario is demonstrated in Fig.~\ref{fig:multirates}.
To avoid the effect of randomness, the results of Fig.~\ref{fig:multirates} are the averages of 100 runs.
Note that, for visualization purposes, LSMR, FPR, and CFMR are magnified in Fig.~\ref{fig:multirates}.
In general, NR has the lowest data rate performance. In Fig.  \ref{fig:multirates}, it is seen that the least and the most data rates belong to FPR and CFMR, respectively.
The data rate of LSMR is 0.53\% greater than that of FPR.
It is found that CFMR results in 2.14\% and 1.61\% more data rates than FPR and LSMR, respectively. 
Since FPR is mobility unaware, it keeps its position; therefore, it  negatively affects the data rate performance of FPR as the victim ships move. 
It is an expected outcome that CFMR and LSMR are better than FPR in terms of data rate performance because of their mobility-aware nature.
There is a small difference between the data rate performances of CFMR and LSMR, which can be related to the area size. Intuitively, as the area size increases, the gap in the data rate performance between CFMR and LSMR widens.

\begin{figure}[h!] 
\centering
\includegraphics[width=1\linewidth]{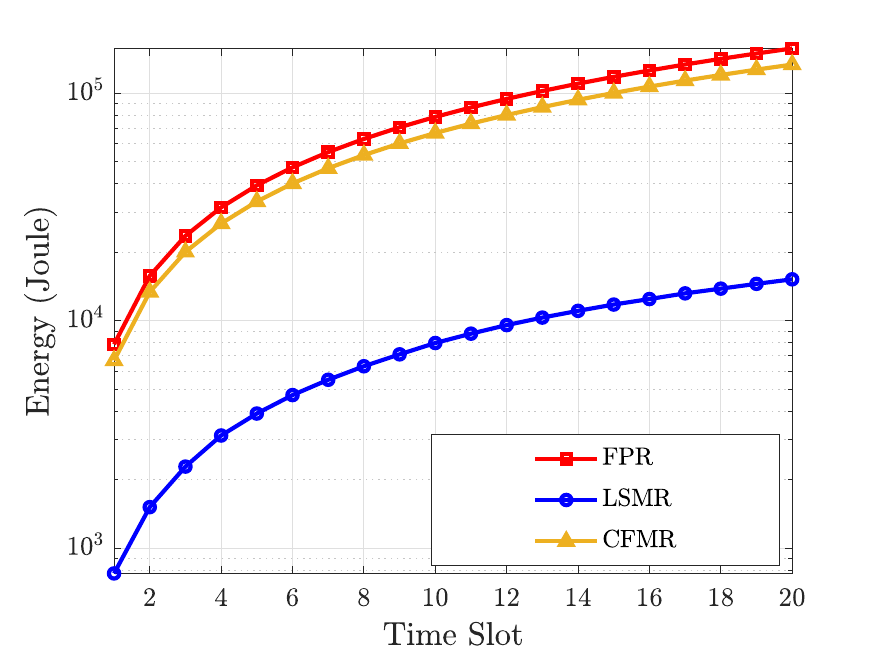}
\caption{Energy consumption for multiple-victim-ship scenario.}
\label{fig:multipleenergy}
\end{figure}

Fig. \ref{fig:multipleenergy} demonstrates the energy consumption performances in the multiple-victim-ship scenario. The energy consumption values in Fig.~\ref{fig:multipleenergy} are cumulative, such that a value at a given time slot is the summation of the energy consumption at that time slot and at all the previous time slots.
As in Fig. \ref{fig:multirates}, the results in Fig. \ref{fig:multipleenergy} are the averages of 100 runs. 
NR is not included because it does not have any UAV relaying mechanism. The highest energy consumption is in FPR, while LSMR results in the least energy consumption.
CFMR yields 15.45\% less energy consumption than FPR, and LSMR reduces the energy consumption of FPR by 90.34\%. 
Furthermore, LSMR achieves 88.58\% less energy consumption than CFMR. 
The reason why LSMR is the best communication architecture in terms of energy consumption is that it has less flight time between LSs. 
FPR also consumes hovering energy over time. On the other hand, in CFMR, the UAV uses less energy due to its speed. 
Intuitively, when the speed of the UAV rises, the mobility energy consumption will soar, in which case CFMR may result in a higher energy consumption than FPR, but LSMR will still be advantageous.

Taking the data rate and energy consumption results into the account together, CFMR yields 1.61\% better data rates than LSMR, but LSMR outperforms CFMR by 88.58\% in terms of energy consumption. 
This navigates us to examine the trade-off between the data rate and energy consumption.
It can be observed that although there is a small difference between LSMR and CFMR in terms of the data rate comparison, there is a huge gap in the energy consumption aspect. 
Hence, the small difference in data rates can be neglected when the huge advantage of LSMR in the energy consumption is considered.

\section{Conclusion}
\label{sec:conclusion}
Smart port technology is a step toward achieving more efficient processes in multiple fields, from transport to finance. Having ubiquitous and reliable wireless connectivity is deemed a fundamental requirement to enable smart ports. However, it is also possible that the harsh harbor environment could render some communication links either fully blocked or underperforming. This issue mainly originates from the shadowing effect, which occurs when there is an obstacle between the transmitter and receiver—quite common in maritime communications. This paper investigates the use of UAV relays to enhance the quality of heavily shadowed or blocked links. Various UAV communication architectures were developed and tested with respect to their data rate and energy consumption performances. Moreover, single- and multiple-victim-ship scenarios were studied independently. In terms of data rate performance, CFMR yielded the best results; however, the minimum energy consumption was obtained with LSMR.      
\bibliographystyle{IEEEtran}
\bibliography{output}

\end{document}